# Determination of the Thermodynamic Scaling Exponent from Static, Ambient-Pressure Quantities

R. Casalini and C.M. Roland

*Naval Research Laboratory, Chemistry Division, Code 6120, Washington, D.C. 20375-5342*

(March 5, 2014)

**Abstract**

An equation is derived that expresses the thermodynamic scaling exponent, $\gamma$, which superposes relaxation times and other measures of molecular mobility determined over a range of temperatures and densities, in terms of static, physical quantities. The latter are available in the literature or can be measured at ambient pressure. We show for 13 materials, both molecular liquids and polymers, that the calculated $\gamma$ are equivalent to the scaling exponents obtained directly by superpositioning. The assumptions of the analysis are that the glass transition is isochronal and that the first Ehrenfest relation is valid; the first assumption is true by definition, while the second has been corroborated for many glass-forming materials at ambient pressure. However, we find that the Ehrenfest relation breaks down at elevated pressure, although this limitation is of no consequence herein, since the appeal of the new equation is its applicability to ambient pressure data.

An important development in understanding the dynamics of supercooled liquids was the discovery that relaxation times, $\tau$, viscosities, $\eta$, diffusion constants, $D$, and other measures of molecular mobility can be expressed as a function of the product of temperature, $T$, and specific volume, $V$, with latter raised to a material constant [1,2,3]. Thus, for the relaxation time

$$\tau = f\left(TV^{\gamma}\right) \tag{1}$$

where $f$ is a function that is unknown *a priori*; similar equations can be written for the other dynamic properties. Eq.(1) has been experimentally validated for more than 100 liquids and polymers, with data from dielectric spectroscopy, dynamic light scattering, and viscosity measurements superposing when plotted versus $TV^{\gamma}$ [4,5]. The only materials exhibiting deviations from eq.(1) are those that undergo changes in chemical structure, such as their degree of hydrogen bonding, with change in state point [6]. Thermodynamic scaling has also been applied to results from molecular dynamics (MD) simulations [7], and eq.(1) is found to be accurate for realistic densities, although for extremes in $V$ some variation in $\gamma$ is necessary for accurate superpositioning [8]. MD simulations have also identified other characteristics of liquids related to thermodynamic scaling [9]. Prominent among these is the strong correlation of fluctuations in the virial pressure ($W$) with fluctuations of the potential energy ($U$); the proportionality constant is equal to the thermodynamic scaling exponent, $\frac{dW}{dU} = \gamma$ [10,11].

One consequence of *dW-dU* correlation is it provides a route to calculation of $\gamma$ from linear thermoviscoelastic response functions, specifically (the dynamic components of) the compressibility, heat capacity, expansion coefficient, and shear modulus [12]. The method obviates the need for measurements at elevated pressures or densities; $\gamma$ can be determined from measurements at one ambient-pressure temperature. This approach has been demonstrated for a silicone oil, DC704, at 214K [12]; the results, $\gamma = 6 \pm 2$, were consistent with the scaling exponent determined from superposition of $\tau(T,V)$, $\gamma = 6.2 \pm 0.2$. The limitation of the method is the difficulty of accurate measurements of frequency-dependent thermoviscoelastic response functions.

A route to $\gamma$ that does not entail any dynamic measurements takes advantage of the fact that the relaxation time at the glass transition is constant (this is true by definition in a kinetic interpretation of vitrification). It follows that the ratio $T_g V_g^\gamma$ is constant, where the subscript refers to the pressure-dependent glass transition. The scaling exponent is determined as [4]

$$\gamma = -\left(\partial \log T_g / \partial \log V_g\right) \qquad (2)$$

Eq.(2) has particular significance in the study of liquid crystals because the bracketed quantity, with the subscript denoting a phase transition (and known as the thermodynamic potential parameter), is central to models of the phase stability of liquid crystals [13]. Conformance of the $\gamma$ from eq.(2) with the value obtained by superpositioning the rotational relaxation times of liquid crystals indicates constancy of the latter at the clearing line (i.e., at state points demarcating the nematic-isotropic phase transition) [14,15].

A general use of thermodynamic scaling is to provide a means to efficiently categorize relaxation data obtained over a broad range of thermodynamic conditions. More significantly, $\gamma$ is a measure of the relative roles of thermal energy and density in governing the dynamics; for example, the scaling exponent can be related to the activation energy ratio [16]

$$E_V / H_P = \left(1 + \alpha_P T_g \gamma\right)^{-1} \qquad (3)$$

in which $E_V = R \left.\dfrac{\partial \ln \tau}{\partial T^{-1}}\right|_\rho$ is the isochoric activation energy and $H_P = R \left.\dfrac{\partial \ln \tau}{\partial T^{-1}}\right|_P$ the activation enthalpy at constant pressure. The advantage of $\gamma$ is that it is a material constant, whereas $E_V/H_P$ varies with state point [4]. An intriguing finding from MD simulations is the connection between the magnitude of $\gamma$ and the steepness of the intermolecular potential in the region around the mean separation distance of Lennard-Jones particles [17,18,19,20]. Of practical utility is that knowledge of $\gamma$ enables $\tau$ (or $\eta$, $D$, …) to be calculated for any thermodynamic condition from measurements at only ambient pressure. However, what has heretofore been lacking is a way to quantify $\gamma$ without the necessity of carrying out experiments at elevated pressure or making relaxation measurements. In this note we describe a procedure to accomplish this.

Eq.(3) can be combined with the Naoki equation [21]

$$\left.\frac{\partial T}{\partial P}\right|_\tau = \left(1 - \frac{E_V}{H_P}\right)\left.\frac{\partial T}{\partial P}\right|_V \tag{4}$$

to yield for the scaling exponent

$$\gamma = \frac{(T_g \kappa_T)^{-1} \left.\frac{\partial T}{\partial P}\right|_\tau}{1 - \left.\frac{\partial T}{\partial P}\right|_\tau \frac{\alpha_P}{\kappa_T}} \tag{5}$$

where $\alpha_P$ and $\kappa_T$ are respectively the thermal expansion coefficient and compressibility. Continuity of the entropy at the glass transition yields [22,23]

$$\left.\frac{\partial T}{\partial P}\right|_\tau = \frac{\Delta \alpha_P T V}{\Delta c_P} \tag{6}$$

in which the $\Delta$'s denote the change at the glass transition. Eq.(6), known as the first Ehrenfest relation, when substituted in eq.(5) gives

$$\gamma = \frac{V \Delta \alpha_P}{\Delta c_P \kappa_T - T V \alpha_P \Delta \alpha_P} \tag{7}$$

Eq. (7), our main result, expresses the scaling exponent in terms of physical units that can be measured at ambient pressure, without relaxation measurements.

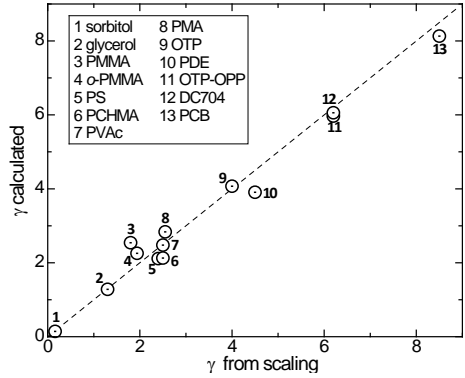

Figure 1. Scaling exponent calculated from eq. (7) versus the value of $\gamma$ obtained by superposition of relaxation times. PMMA: polymethymethacrylate (high polymer); o-PMMA: oligomeric PMMA; PS: polystyrene; PCHMA: polycyclohexylmethacrylate; PVAC: polyvinylacetate; PMA: polymethacrylate; OTP: ortho-terphenyl; PED: phenolphthalein dimethylether; DC704: tetramethyl tetraphenyl trisiloxane; OTP-OPP: mixture of 67% OTP and 33% ortho-phenylphenol; PCB: chlorinated biphenyl. The dashed line represents equivalence of the $\gamma$. References for the data are in Table 1.

There are two assumptions underlying eq.(7): at $T_g$ $\tau$ is constant and the entropy is continuous. As stated, the first assumption is true by definition, although that fact that the glass transition is a kinetic phenomenon means the values of some parameters in eq.(7) are sensitive to thermal and pressure histories; this potentially introduces uncertainty into the calculated $\gamma$.

The second assumption has been verified, at least at ambient pressure, for many glass-formers [24- 29].

We test eq.(7) directly by comparing the computed $\gamma$ to values obtained in the usual fashion by superposition of relaxation times. This is done for thirteen liquids for which the thermodynamic data in eq.(7) are available (Table 1) [12,30- 43]. The results, shown in Figure 1, affirm the correctness of the new expression for $\gamma$.

Although the present analysis allows $\gamma$ to be determined from ambient-pressure data, the quantities in eq.(7) can also be measured at elevated pressures. For normal liquids ("correlating liquids" in the parlance of ref. [44]) $\gamma$ is constant, so the results would be the same. However, for liquids with structure that changes with temperature or pressure, such as the concentration of H-bonds, $\gamma$ is expected to vary. Eq.(7) cannot be applied in such situations because of the limited validity of eq.(6) at high pressures. This is illustrated in Figure 2 for polyvinylacetate using the equations of state for the liquid and glass from [38]. The pressure dependence of $\Delta c_p$ was calculated assuming the heat capacity of the glass is constant and that of the liquid varies as $\int_0^P T \frac{\partial V^2}{\partial T^2} dP'$. Although Eq.(6) is accurate at atmospheric pressure, it underestimates $\partial T/\partial P|_\tau$ at higher pressures. (This deviation of Eq. (6) at elevated $P$ appears to be a new observation.) We also found that for sorbitol (data not shown) the departure of $\partial T/\partial P|_\tau$ from experimental data was as much as 50% at 100MPa.

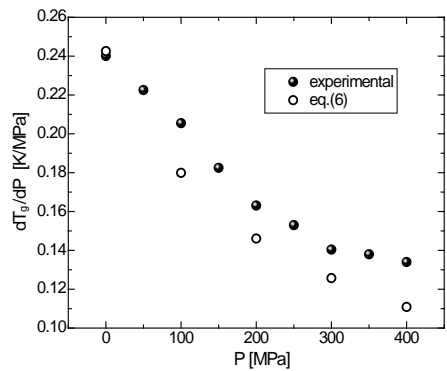

Figure 2. Thermal pressure coefficient at $T_g$ as a function of pressure determined experimentally (filled symbols) and calculated using the first Ehrenfest relation (open symbols). The latter's validity is limited to ambient pressure. Scaling exponent calculated from eq. (7) versus the value of $\gamma$ obtained by superposition of relaxation times. PMMA: polymethymethacrylate (high polymer); o-PMMA: oligomeric PMMA; PS: polystyrene; PCHMA: polycyclohexylmethacrylate; PVAC: polyvinylacetate; PMA: polymethacrylate; OTP: ortho-terphenyl; PED: phenolphthalein dimethylether; DC704: tetramethyl tetraphenyl trisiloxane; OTP-OPP: mixture of 67% OTP and 33% ortho-phenylphenol; PCB: chlorinated biphenyl. The dashed line represents equivalence of the $\gamma$. References for the data are in Table 1.

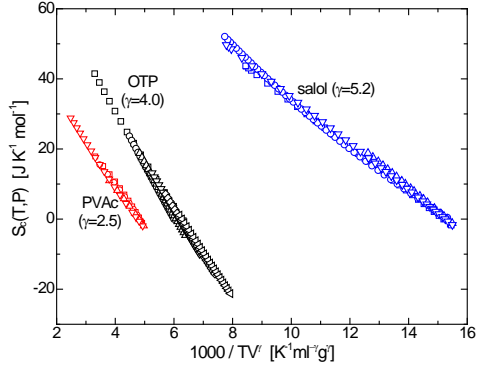

Figure 3. Configurational entropy of three liquids measured at various temperatures and pressures plotted versus the scaling variable with the indicated value of γ. Data from ref. [46,47].

We previously showed [39] that the thermodynamic scaling property can be derived from models that connect the supercooled dynamics to the entropy [45]. These also provide an alternative route to eq.(7). To show this we note that the configurational entropy, $S_C$, conforms to thermodynamic scaling [46,47] (Figure 3)

$$S_c = g(TV^\gamma) \qquad (8)$$

where $g$ is a function, and the scaling exponent is the same $\gamma$ as in eq.(1). From the continuity of the entropy at the glass transition the derivatives of temperature and volume are related as

$$\left(\left.\frac{\partial S^{liquid}}{\partial T}\right|_V - \left.\frac{\partial S^{glass}}{\partial T}\right|_V\right)dT = -\left(\left.\frac{\partial S^{liquid}}{\partial V}\right|_T - \left.\frac{\partial S^{glass}}{\partial V}\right|_T\right)dV \qquad (9)$$

Since the non-configurational component of the entropy is unaffected by vitrification [39,45], it cancels out and eq.(9) can be rewritten as

$$\left(\left.\frac{\partial S_C^{liquid}}{\partial T}\right|_V - \left.\frac{\partial S_C^{glass}}{\partial T}\right|_V\right)dT = -\left(\left.\frac{\partial S_C^{liquid}}{\partial V}\right|_T - \left.\frac{\partial S_C^{glass}}{\partial V}\right|_T\right)dV \qquad (10)$$

It follows from the scaling property of $S_C$ (eq.(8))

$$\left.\frac{\partial S_{conf}}{\partial V}\right|_T = \gamma \frac{T}{V} \left.\frac{\partial S_{conf}}{\partial T}\right|_V \qquad (11)$$

Eq. (11) bears a formal similarity to the relation between the Grüneisen parameter and the derivatives of $S$ [39]. Combining eqs. (10) and (11) yields

$$\left.\frac{\partial V}{\partial T}\right|_\tau = -\frac{V}{T\gamma} \qquad (12)$$

Rewriting the temperature derivative of the volume at constant $\tau$ as

$$\left.\frac{\partial V}{\partial T}\right|_\tau = \left.\frac{\partial V}{\partial T}\right|_P + \left.\frac{\partial V}{\partial P}\right|_T \left.\frac{\partial P}{\partial T}\right|_\tau \tag{13}$$

it can be seen that eq.(12) is equivalent to eq.(5), and thus eq.(7) is obtained.

Table 1. Physical quantities used to calculate the scaling exponent.

| $\left.\frac{\partial T}{\partial P}\right|_\tau$ ¥ [KMPa$^{-1}$] | $\alpha_P^{liq} \times 10^4$ [K$^{-1}$] | $\Delta\alpha_P \times 10^4$ [K$^{-1}$] | $\Delta c_P$ [Jmol$^{-1}$K$^{-1}$]* | $V$ [cm$^3$mol$^{-1}$] | $\kappa_T \times 10^4$ [MPa$^{-1}$] | $T_g$ [K] | $\left.\frac{\partial T}{\partial P}\right|_\tau$ † [KMPa$^{-1}$] | $\gamma$ | $\gamma_{exp}$ | ref. |
|---|---|---|---|---|---|---|---|---|---|---|
| sorbitol | 4.46 | 2.74 | 189.4 | 111.7 | 11.5 | 272 | 0.044 | 0.14±0.012 | 0.16 | 30 |
| glycerol | 4.8 | 2.4 | 81 | 69.99 | 1.8 | 183 | 0.0379 | 1.28±0.15 | 1-1.6 | 31 |
| PMMA | 5.24 | 2.9 | 37.2 | 86.96# | 3.9 | 380 | 0.258 | 2.8±0.34 | 1.8 | 32 |
| o-PMMA | 6.2 | 3.8 | 43.1 | 84.22 | 4.85 | 338.9 | 0.251 | 2.3±0.2 | 1.94 | 33,34 |
| PS | 6.0 | 3.2 | 34 | 100.8 | 6.5 | 353 | 0.328 | 2.1±0.3 | 2.5 | 35,36 |
| PCHMA | 5.36 | 1.5 | 33.6 | 161.5 | 4.7 | 336 | 0.245 | 2.1±0.4 | 2.5 | 37 |
| PVAc | 7.15 | 4.52 | 40.7 | 72.5 | 5.0 | 304 | 0.245 | 2.48±0.14 | 2.5 | 38,39 |
| PMA | 6.64 | 3.34# | 37.8 | 70.28 | 3.8 | 287 | 0.201 | 2.83±0.3 | 2.55 | 40 |
| OTP | 7.08 | 5.49 | 113 | 205.9 | 4.2 | 246 | 0.246 | 4.05±0.3 | 4 | 41 |
| PDE | 6.08 | 3.16 | 96.8 | 255.07 | 3.64 | 298 | 0.248 | 3.91±0.4 | 4.5 | 39,42 |
| OTP-OPP | 8.5 | 5.6 | 123 | 203.9 | 3.4 | 233.7 | 0.217 | 6.0±0.3 | 6.2 | 35 |
| DC704 | 4.6 | 3.5 | 145.4 | 425.18 | 2.5 | 212 | 0.217 | 6.8±0.8 | 6.2 | 12 |
| PCB | 7.50 | 3.4 | 66.1 | 239.6 | 4.0 | 268.9 | 0.330 | 8.1±0.86 | 8.5 | 43 |

*per repeat unit for polymers; # $\alpha_P^{glass}$ taken at 200MPa ; † calculated using eq.(6).

Since the discovery of thermodynamic scaling [1,2,3] much experimental and theoretical work has been done to relate the exponent $\gamma$ to physical properties of the materials [4,5], such as activation energies for the dynamics [16] and the steepness of the intermolecular potential [17,18,19,20]. There also exist expressions relating certain thermodynamic properties and their changes at the liquid-glass transition. An example of the latter is the first Ehrenfest equation, which has been corroborated for a large number of materials [24,25,26,27,28,29]. Combining this equation with a formula for $\gamma$, we derive an expression for the scaling exponent in terms of static, physical properties. This expression also follows from thermodynamic scaling of the configurational entropy. The predictions of this new relation are found to be in agreement with results obtained directly by superposition of relaxation data. This analysis enables dynamics properties to be determined for any thermodynamic condition from quantities that are routinely measured or available in the open literature. The fact that the accuracy of the Ehrenfest equation appears to be limited to low pressures means that the analysis can only be implemented using ambient pressure quantities; however, this limitation is inconsequential, since the advantage of the method is obviating the requirement for high pressure measurements.

**Acknowledgement**

This work was supported by the Office of Naval Research.